\documentclass{aastex631}

\begin{document}

\title{A 50-min coronal kink oscillation and its possible photospheric counterpart} 


\correspondingauthor{Valery M. Nakariakov}
\email{v.nakariakov@warwick.ac.uk}

\author[0000-0002-5606-0411]{Sihui Zhong}
\affiliation{Centre for mathematical Plasma Astrophysics, Department of Mathematics, KU Leuven,\\
Leuven, BE-3001, Belgium}
\affiliation{Engineering Research Institute \lq\lq Ventspils International Radio Astronomy Centre (VIRAC)\rq\rq\ of Ventspils University of Applied Sciences,\\ Inzenieru iela 101, Ventspils, LV-3601, Latvia}

\author[0000-0001-6423-8286]{Valery M. Nakariakov}
\affiliation{Centre for Fusion, Space and Astrophysics, Department of Physics, University of Warwick,\\ Coventry CV4 7AL, UK}
\affiliation{Centro de Investigacion en Astronom\'ia, Universidad Bernardo O'Higgins, Avenida Viel 1497, Santiago, Chile}

\author[0000-0002-0687-6172]{Dmitrii Y. Kolotkov}
\affiliation{Centre for Fusion, Space and Astrophysics, Department of Physics, University of Warwick,\\ Coventry CV4 7AL, UK}
\affiliation{Engineering Research Institute \lq\lq Ventspils International Radio Astronomy Centre (VIRAC)\rq\rq\ of Ventspils University of Applied Sciences,\\ Inzenieru iela 101, Ventspils, LV-3601, Latvia}

\begin{abstract}
A coronal loop of 290~Mm length, observed at 171~\AA\ with SDO/AIA on February 6th 2024 near AR 13571, is found to oscillate with two significantly different oscillation periods, $48.8 \pm 6.1$~min and $4.8\pm 0.3$~min. The oscillations occur in the time intervals without detected flares or eruptions. Simultaneously, near the Northern footpoint of the oscillating loop, we detect a $49.6 \pm 5.0$-min periodic variation of the average projected photospheric magnetic field observed with SDO/HMI. The shorter-period decayless oscillation is attributed to the eigen-mode, standing kink oscillation of the loop, while the longer-period oscillation may be the oscillatory motion caused by the periodic footpoint driver. The photospheric long-period process can also drive the short-period, eigen oscillation of the loop via the self-oscillatory, \lq\lq violin\rq\rq\, mechanism, in which a transverse oscillation is excited by an external quasi-steady flow. This finding indicates that the most powerful, lower-frequency spectral components of photospheric motions, which are well below the Alfv\'enic/kink cutoff, can reach the corona. 


\end{abstract}

\keywords{Solar Corona (1483) --- Solar coronal waves(1995) --- Solar coronal loops(1485) --- Solar Photosphere (1518)}

\section{Introduction} \label{sec:intro}

The problem of heating the solar and stellar coronae to temperatures exceeding one million degrees remains one of the long-standing puzzles in modern Astrophysics. 
It is commonly accepted that the energy required for coronal heating comes from dynamic processes in the lower atmosphere \citep{2015RSPTA.37340256K, 2015RSPTA.37340269D,2020SSRv..216..140V}.
The power spectra of photospheric dynamics are typically characterized by a power-law dependence \citep[e.g.,][]{2010ApJ...716L..19M,2010ApJ...710.1857M,2012ApJ...752...48C,2016A&A...592A.153K}, that is, the spectral power increases as the frequency decreases. The power spectrum of coronal motions also exhibits a power-law shape \citep{2025ApJ...986L...6M}. 
However, it remains unclear whether the most energetic, low-frequency components of photospheric motions can reach the corona, as establishing a direct connection between photospheric random motions and those in the corona is challenging. This association could potentially be inferred through the sporadic detection of narrowband low-frequency oscillatory processes, which might serve as indicators of the energy transfer of powerful broadband low-frequency motions from the photosphere to the corona.

There have been occasional detections of photospheric oscillations with periods from 10~min to hours, i.e., with periods longer than the acoustic cutoff period, $\sim$5~min, in radio \citep{1972SoPh...25..339K, 2006PASJ...58...29G, 2007ARep...51..401E, 2009CosRe..47..279E, 2010A&A...513A..27C, 2013PASJ...65S..12A} and optical \citep{2007ARep...51..401E,2017A&A...598L...2K, 2017A&A...597A..93D, 2020A&A...635A..64G, 2021A&A...654A..50N, 2022SoPh..297..154C} bands, and also in those bands simultaneously \citep{2013A&A...552A..23S}.
An incomplete list of the mechanisms which are possibly responsible for these oscillations includes supergranular forcing \citep{2016Ge&Ae..56.1052S}; vortex shedding induced by flux emergence \citep{2001ApJ...549.1212E}; granular buffeting \citep{2013A&A...559A..88S, 2013A&A...554A.115S}; vortex motions \citep{2010ApJ...723L.139B}; the buoyancy oscillations, i.e. the g(gravity)-mode \citep{2010A&ARv..18..197A}, guided to the solar surface by magnetic flux tubes \citep{2011A&A...533A.116Y}; and natural global oscillations of sunspots \citep{2008AstBu..63..169S, 2021A&A...653A..39D}.  

An abundant number of magnetohydrodynamic (MHD) wave phenomena have also been observed in the corona \citep[e.g.,][]{2014SoPh..289.3233L,2020ARA&A..58..441N, 2020SSRv..216..136L, 2021SSRv..217...76B, 2021SSRv..217...34W}. One of the most studied coronal oscillatory processes is kink oscillations characterized by a transverse displacement of coronal loops or other plasma structures \citep[e.g.,][]{2021SSRv..217...73N}. The oscillation periods are typical of coronal origin, being consistent with the kink resonance in waveguiding coronal plasma structures \citep{2019ApJS..241...31N, 2021SSRv..217...73N, 2023NatSR..1312963Z}.  

A simultaneous detection of 29--75~min oscillations in the corona in EUV and the upper chromosphere in microwaves has been reported before \citep{2011A&A...533A.116Y}. However, the EUV signal was contaminated with the spacecraft's orbital period artefacts, and hence that finding is not reliable. Thus, so far, there has been no direct reliable evidence for low-frequency energy being tunneled from lower atmospheric layers to the corona. 

In this letter, we present a simultaneous detection of a 50-min oscillation of the line-of-sight (LoS) component of the photospheric magnetic field with the Helioseismic and Magnetic Imager (HMI, \citealt{2012SoPh..275..207S}) on board the Solar Dynamics Observatory (SDO, \citealt{2012SoPh..275....3P}), and a 50-min transverse oscillation of a coronal plasma loop observed with the Atmospheric Imaging Assembly (AIA, \citealt{2012SoPh..275...17L}). Our finding suggests that low-frequency photospheric motions could reach the corona and drive coronal loop oscillations.

\section{Observational data processing and analysis \label{sec:methods}}
On February 6, 2024, between 12:00 and 17:00 UT, SDO/AIA captured kink oscillations in several high-contrast coronal loops near active region (AR) 13571, located about the solar disk center. At the same time, oscillations in the line-of-sight magnetic field (B$_\mathrm{LoS}$) were observed at one of the loop footpoints with SDO/HMI.

In this work, SDO data, including AIA images at multiple wavelengths, HMI LoS magnetograms, HMI intensity continuum, and HMI Dopplergrams, are used to analyze the observed coronal and photospheric oscillations. Before downloading from the Joint Science Operations Center (JSOC), data are processed with subframe extracting, tracking, and registering, with the latter two aiming at solar de-rotation.
The AIA images have a time cadence of 12\,s and a pixel size of 0.6\,arcsec. Among all bandpasses of AIA, 171\,\AA\ images with a formation temperature peak at around 1~MK are best to detect high-contrast coronal features, so they are commonly used to analyze oscillations in the coronal loops. In addition, images in 1600\,\AA\ and 1700\,\AA\ are looked at in the search for the oscillatory low-atmospheric activity.
The HMI LoS magnetograms and the intensity continuum have a spatial resolution of 1\,arcsec and a time cadence of 45~s. The accuracy of HMI B$_{\rm LoS}$ in one pixel is 12--15~G, and the error of average B$_{\rm LoS}$ over a box with $n\times n$ pixels is $\sqrt{n^2}/n^{2}\times15$~G, according to the error propagation formula. For a region of interest (ROI) with $73\times 73$ pixels,  the error of B$_{\rm LoS}$ is 0.2~G.


The analysis begins with processing AIA 171\,\AA\ images using the wavelet-optimized whitening (WOW) technique \citep{2023A&A...670A..66A} to enhance contrast and suppress noise. We adopt the original image weight of 0.995 ($gw=0.995$), and denoising thresholds of 10 and 5 for the first two wavelet scales ($d=10,5$). Transverse loop oscillations are then detected with the help of the motion magnification technique \citep{2016SoPh..291.3251A,2021SoPh..296..135Z} with a magnification factor of 8. This method linearly amplifies subtle transverse displacements in image sequences to visible levels.

To extract oscillation signals, time--distance maps are constructed using slits placed perpendicular to the loop axis, each 5 pixels wide, to improve the signal-to-noise ratio via intensity averaging. At each time step, the transverse intensity profile is fitted with a Gaussian to determine the loop center, generating a time series of transverse displacements.
The signals are detrended using empirical mode decomposition \citep[EMD,][]{1998RSPSA.454..903H,2008RvGeo..46.2006H}: the last intrinsic mode function (IMF) and residual from two EMD iterations are summed and subtracted from the original signal to isolate the oscillatory component. The oscillation amplitude and its uncertainty are calculated as the mean and standard deviation of instantaneous amplitudes, calculated via the Hilbert transform of the detrended signal.

Periodicity is then estimated using, independently, Fast Fourier Transform (FFT), wavelet analysis \citep{1998BAMS...79...61T}, and EMD \citep{2016A&A...592A.153K, 2022SSRv..218....9A}. In this work,  all periodicity analyses are performed with a red-noise significance test to distinguish oscillatory signals from noise.

Moreover, the H$\alpha$ Imaging Spectrograph (HIS) on board the Chinese H$\alpha$ Solar Explorer (CHASE; \citealt{2022SCPMA..6589602L}) are employed to obtain the chromospheric Doppler velocity field. HIS scanned full-disk H$\alpha$ spectra for 20~min for each hour from 12:00 to 19:00 UT on February 6 2024. For each data sequence, the temporal resolution is 70~s, the spatial resolution is 1.2~arcsec, and the spectral resolution is 0.048\,\AA. CHASE data are available in the Solar Science Data Center of Nanjing University (SSDC) \footnote{\url{https://ssdc.nju.edu.cn}}. To obtain the chromospheric Doppler velocity distribution, we first extract a reference spectral H$\alpha$ line by averaging spectra in a quiet Sun region of $100\times100$ pixels. Then the Doppler velocity in each pixel is determined by the shift between the analyzed spectral line and the reference. The CHASE orbital velocity and other artificial Doppler shifts caused by the differential rotation, meridional flows, and limb shift are subtracted to get the intrinsic velocity \citep[see methods in][]{2024NatAs...8.1102R,2024ApJ...968L..20Y}.

\section{Results \label{sec:results}}
\begin{figure*}
	\centering
        \includegraphics[width=\linewidth]{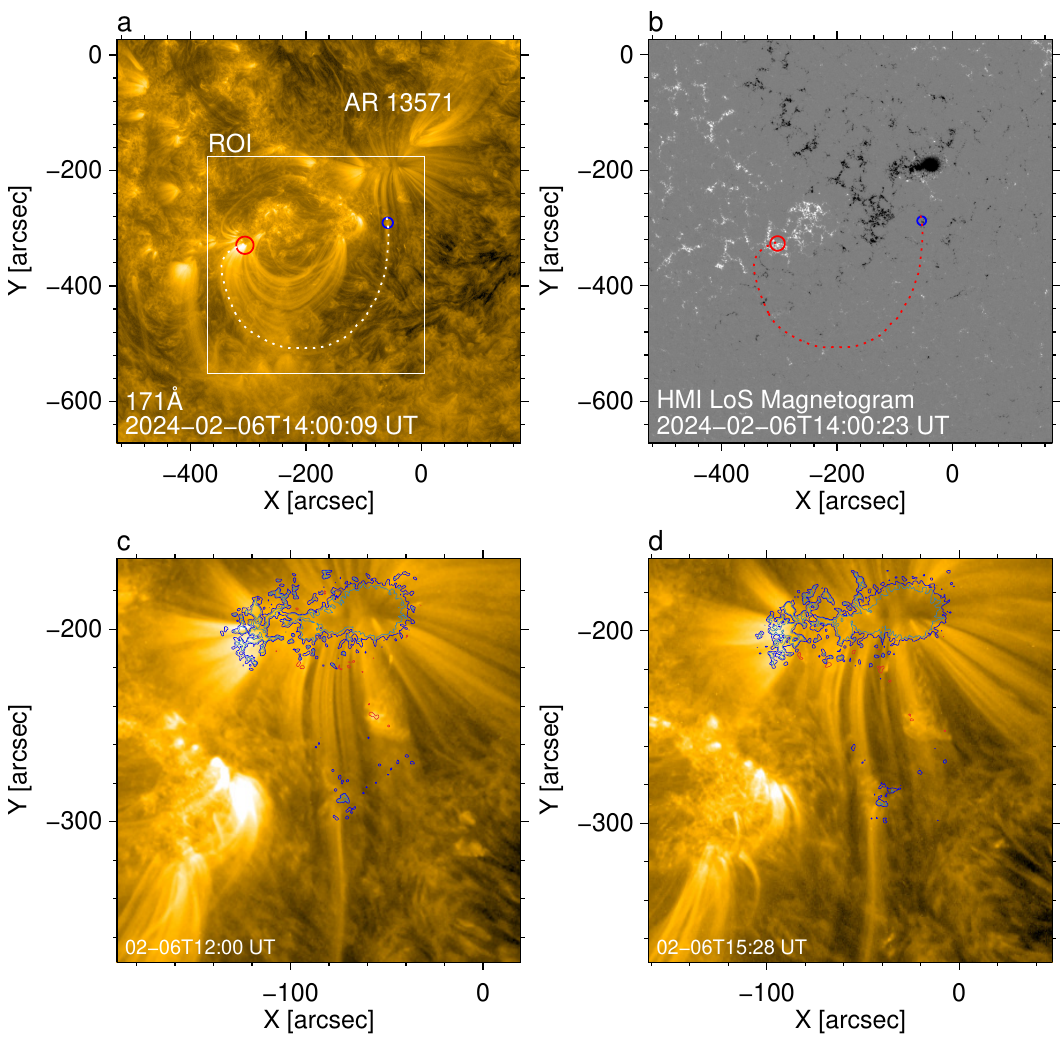}
	\caption{
        The region of study that displays the oscillating loops by AIA 171 \,\AA\ image (a) and corresponding HMI LoS magnetogram (b). The box in panel~a outlines the region of interest (ROI). The dotted curves sketch the loop of interest, and the red/blue circles denote their positive/negative footpoints. Panels~(c--d) show the magnetic connectivity of the western footpoint of the analyzed loop. Blue/light blue/red contours indicate the magnetic field of -250/-100/250~G. An animation of panels (c) and (d) showing the temporal evolution of the Western footpoint of the studied coronal loop is available. The animation begins at 12:00 UT and ends at 16:59 UT. The real-time duration of the animation is 12 seconds.
	}
	\label{fig:FOV}
\end{figure*}

On February 6th 2024 from 12:00~UT to 17:00~UT, AIA in the 171\,\AA\ channel detected kink oscillations in several high-contrasted coronal loops near AR~13571. Figure~\ref{fig:FOV} displays the ROI in the AIA 171\,\AA\ image and HMI LoS magnetogram at around 14:00~UT. In the ROI, two sets of loops exhibit oscillations: one of a shorter loop length of about 223~Mm, and the other of a longer length of 290~Mm (indicated by the dotted curve) and a lifetime of more than 2 days, surviving from 5th to 8th February. Here, we only focus on the kink oscillations in the 290-Mm-long loop.
As shown in Figure~\ref{fig:FOV}b, the loop of interest is anchored in two opposite magnetic patches, and hence is situated over a polarity inversion line. The target loop appears to overlap with the loop fan originating from the sunspot 13571. However, a closer view of the western footpoint region in Figure~\ref{fig:FOV}c--d reveals that the loop is connected to a negative polarity region (also denoted by the smaller blue circle in Figure~\ref{fig:FOV}a--b) located south of the sunspot. This interpretation is further supported by the fact that the loop’s segment near the footpoint is both brighter than the northern loop fan and exhibits an S-shaped structure (see Figure~\ref{fig:FOV}d). A supplementary animation features the evolution of this region.
The length of the loop connecting the positive patches and the negative pore is 290~Mm. The connectivity seems to vary after the kink oscillation, which could be due to a projection effect when observing several overlapping structures along the same LoS. 

\subsection{Coronal loop oscillations} \label{sec:kink}

\begin{figure*}
	\centering
	\includegraphics[width=\linewidth]{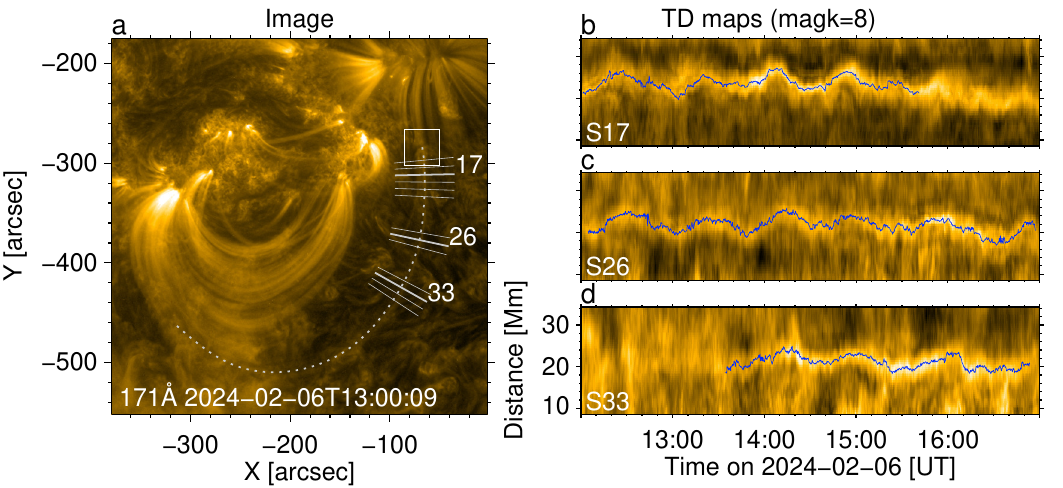}
	\caption{Kink oscillations in the analyzed loop bundle. (a) AIA 171\,\AA\ image showing the loop ensemble marked by the dotted curve. White slits across the loop are used to make time--distance maps that reveal the oscillatory signals. The white box locates the possible negative footpoint of the oscillating loop. (b-d) Representative time--distance maps exhibiting the decayless kink oscillations. The data is magnified by a factor (magk) of 8. The blue curves are oscillation signals extracted by the Gaussian fitting of the transverse intensity profiles.
	}
	\label{fig:50min_td}
\end{figure*}

The loop of interest shows a transverse oscillation that lasts for about 6 cycles (5 hours) without significant damping, see Figure~\ref{fig:50min_td}, which is characteristic of decayless kink oscillations \citep{2012ApJ...751L..27W, 2013A&A...560A.107A, 2015A&A...583A.136A, 2023NatSR..1312963Z}. 
In the time--distance maps, these low-amplitude oscillations are revealed with the help of the motion magnification technique with the magnification factor of 8. 
In Figure~\ref{fig:50min_td}b--d, we show three good examples of the oscillatory patterns determined for three different slits across the loop. The oscillatory signals exhibit two periodicities: a long one of around 50~min, and a shorter one of several minutes. The shorter periodicity appears in the time interval from 14:00 to 15:00~UT, see panel d as a particular example.
The amplitude of the long-period oscillation magnified by a factor of 8 is $3.7\pm1.6-4.8\pm2.3$~Mm, so the original amplitude is $0.4\pm0.2-0.6\pm0.3$~Mm. For short-period signals, the magnified and original amplitudes are approximately $1.2\pm 0.7$~ Mm and $0.15\pm 0.09$~ Mm, respectively. The errors are estimated as standard deviations of the instantaneous amplitudes of the detrended signals.

\begin{figure*}
	\centering
	\includegraphics[width=\linewidth]{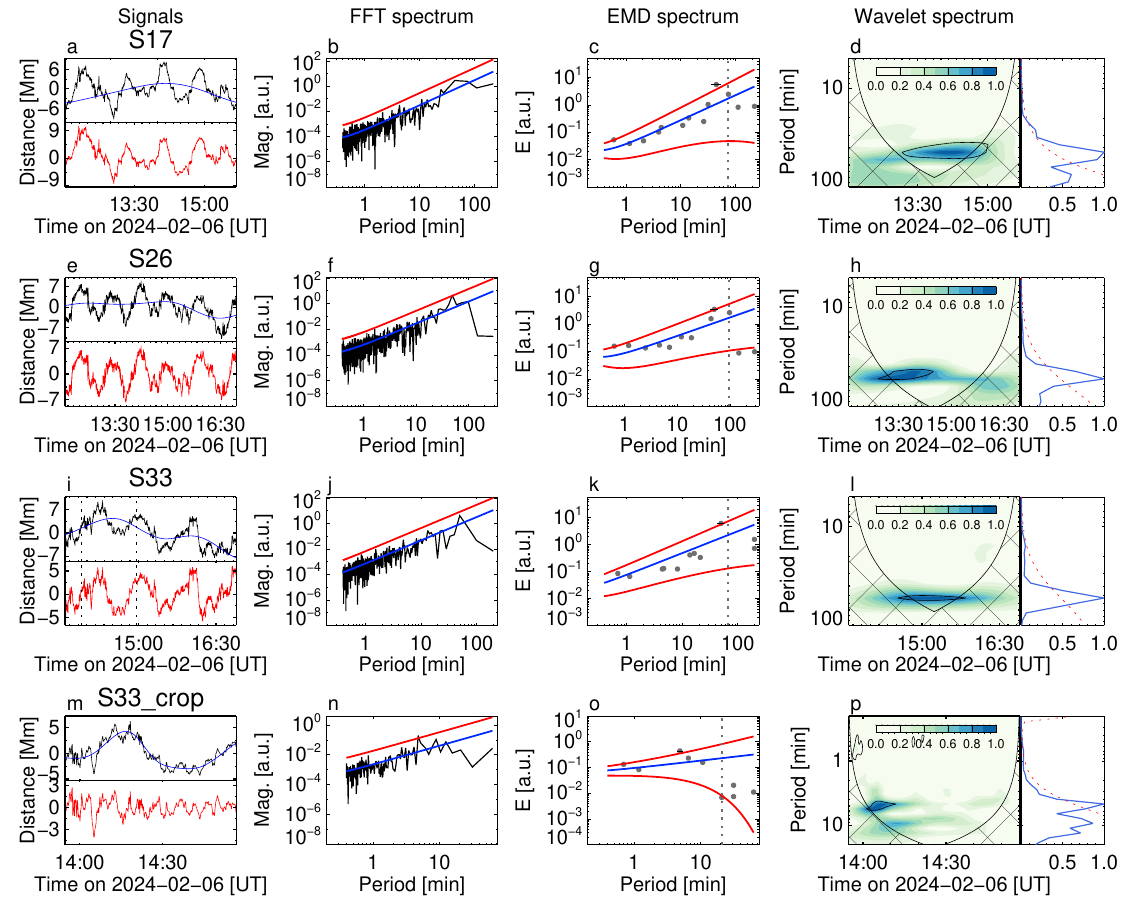}
	\caption{Periodicity analysis of the oscillations in the loop of interest. From top to bottom, each row is for oscillations in Slit 17, 26, 33 (blue curves in Figure~\ref{fig:50min_td}b--d) and selected time interval [13:55~UT, 14:58~UT] 
    in S33. First column: Original signals (black) with their trend (blue) in the upper sub-panel and detrended signals (red) in the bottom sub-panel. The relative variation of the signals is linearly magnified by a factor of 8. From the second to the fourth column are the Fourier power spectrum with a 95\% confidence level (red line), EMD energy spectrum with a 95\% confidence interval (red lines), and Wavelet spectrum with a 95\% confidence level (black contour), respectively. In panel (i), the vertical lines indicate the time interval selected for further detection of shorter periodicity (5 minutes) shown in panels~(m--p). In panels~(b--n), the blue lines are the best-fitting power-law functions. In panels~(c--o), the vertical lines indicate the threshold of reliable period estimation, the blue lines are the mean energy, and grey horizontal lines indicate error bars. In panels~(d--p), the color bars indicate the normalized spectral power, the blue curves in the right subpanels are the global normalized wavelet power as a function of period, and the red dotted curves indicate the 95\% confidence levels. Here, \lq\lq Mag.\rq\rq, \lq\lq E\rq\rq, \lq\lq a.u.\rq\rq\ stands for power magnitude, EMD modal energy and arbitrary units, respectively.
	}
	\label{fig:50min_period}
\end{figure*}

Figure~\ref{fig:50min_period} presents the periodicity analysis of the oscillations. 
The oscillation periods are estimated by three independent methods: Fourier analysis, EMD, and wavelet analysis. We consider the detection of the oscillation to be valid only if the same periodicity is confirmed by at least two different methods. For signals in slit S17, as shown in Figure~\ref{fig:50min_period}b--d, the Fourier power spectrum does not show a periodicity above the 95\% confidence level (red), while in the EMD energy spectrum, the tenth mode with periodicity $44.4\pm9.0$~min is outside the confidence interval 95\% (red), and the Morlet wavelet spectrum shows a peak of $44.5\pm8.5$~min above the confidence level 95\% (red). For S26, the Fourier spectrum shows a peak at $48.8\pm6.1$~min, EMD in the ninth mode at $50.5\pm7.7$~min, and wavelet at $52.9\pm6.7$~min. For S33, no periodicity is detected in the Fourier spectrum, but the ninth EMD mode of $50.1\pm6.9$~min and the $52.9\pm7.6$~min peak at the wavelet spectrum appear.
To summarise, all three signals share a periodicity of around 50~min, with the average value of $49.2\pm7.5$. The failure of periodicity detection in Fourier spectra could be caused by the short duration of the oscillatory signals, i.e., the spectral power is distributed over several neighbouring frequencies \citep[see][]{2017A&A...602A..47P}.
In addition, in the time interval from 13:55 to 14:58~UT, the signals of S33 show a 5-min periodicity: a significant peak of 95\% appears at $4.8\pm0.3$~min in the Fourier power spectrum, the third mode of $4.9\pm0.5$~min in the EMD spectrum, and $4.7\pm0.5$~min in the wavelet spectrum. The average period is $4.8\pm0.4$~min.

Cross-correlation analysis is applied to oscillatory signals in different loop segments to check whether the detected 50-min kink oscillations are propagating or standing. The Pearson cross-correlation coefficient between the signals in the slits S26 and S33, separated by 40.5~Mm is maximum (0.94) with zero delay. The cross-correlation coefficient between S17 and S33 peaks at 0.88 with a time lag of 432~s, which is only 0.14 of the oscillation period. The distance between the slits is 72~Mm. If the observed kink oscillations are propagating, the speed is $72[\mathrm{Mm}]/432[\mathrm{s}] = 167 [\mathrm{km/s}]$, which is an order smaller than the typical value of the kink speed, about 1000~km/s \citep{2021SSRv..217...73N}. Hence, we attribute the small phase shift between the oscillatory signals at S17 and S33 to the effect of noise or maybe dissipation processes and conclude that the detected 50-min oscillation is a standing kink oscillation.

However, for a 290-Mm-long loop oscillating in the normal mode, its period in fundamental harmonic should be about $2\times290 \mathrm{[Mm]}/1000 \mathrm{[km/s]} \approx 580 \mathrm{[s]} \approx 10 \mathrm{[min]}$. 
Those values are several times smaller than the detected 50-min period. The phase speed required for the detected period is about 200~km/s, which is about the sound speed. It corresponds to an unrealistically high value of the plasma parameter $\beta$. The decayless nature of the oscillation suggests the presence of a continuous external driver, possibly with the same periodicity. 
On the other hand, the 5-min oscillation of this loop would require the kink speed to be either about 2000~km/s, if it is the fundamental natural harmonic, or 1000~km/s if it is the second harmonic. The latter value is consistent with typical coronal kink speeds. 


\subsection{Photospheric oscillation} \label{sec:photosphere}
The 50-min periodicity is readily found in the temporal variation of the LoS component of the photospheric magnetic field observed with HMI. The oscillation is detected in the region near the western footpoint of the oscillating coronal loop. 
This region of interest, labelled ROI$_1$, in the HMI LoS magnetogram, is displayed in Figure~\ref{fig:meanB}a, and outlined by the white box in Figure~\ref{fig:50min_td}a.

The time variation of the average projected magnetic field (B$_{\rm LoS}$) in ROI$_1$ from 10:00~UT to 19:00~UT is displayed in Figure~\ref{fig:meanB}b. The 50-min coronal oscillation is shown between the two red vertical lines, while a clear oscillatory pattern of the LoS magnetic field from 12:45 to 17:00~UT is shown between the two blue vertical lines. These two intervals overlap.
The detrended signals are displayed as the black curves in panel c. The variation of the signals is around 1~Gauss (G), which is greater than the error of the average B$_{\rm LoS}$ over ROI$_{1}$ ($73\times73$~pixels), 0.2~G. In the wavelet spectrum of the signals selected from 12:45~UT to 19:00~UT, see panel d, a periodicity around 50~min is within the 95\% confidence contour. The global wavelet spectrum peaks at $49.6\pm5.0$~min, over the 95\% confidence level. In panel e, the sixth mode of the EMD spectrum has the period of $50.3\pm11.2$~min is above the 95\% confidence level, confirming the existence of 50-min periodicity. Mode 6 is overplotted as the red curve in panel c, which is consistent with the detrended signal (black). As shown in the Hilbert spectrum of mode 6 (panel f), the instantaneous period of mode 6 varies around 50~min in time. 

The identification of oscillation mode of the detected photospheric oscillation is attempted by tracking the motion of the loop footpoint in ROI$_1$. As the loop footpoint is rooted in a cluster of negative magnetic pores which are discrete and irregular, we quantify the location of the footpoint by the centroid position of B$_{\rm LoS}$ in ROI$_1$, see Figure~\ref{fig:meanB}g. The equivalent loop footpoint is moving westwards from 12:00~UT (dark purple) to 17:00~UT (yellow), without a transverse (kink-like) trajectory, indicating that the oscillation is not a pure kink mode.

\begin{figure*}
	\centering
	\includegraphics[width=\linewidth]{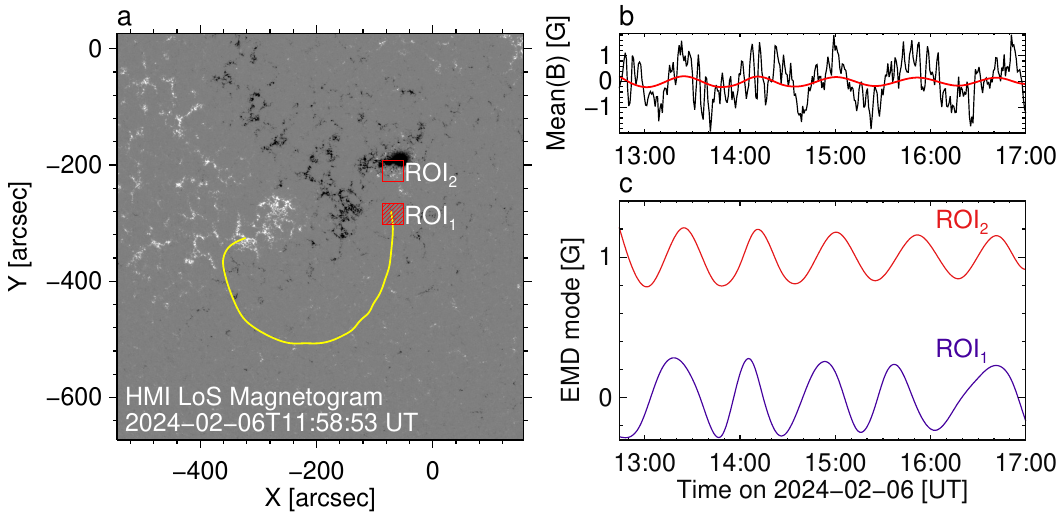}
	\caption{The spatial location of 50-min oscillation of the mean magnetic field in the study region. (a) HMI LoS magnetogram overlaid with 2 boxes outlining the areas exhibiting 50-min periodicity. The yellow spline depicts the oscillating loop. The filled area (labelled ROI$_{1}$) covers the footpoint area of the loop. (b) The detrended mean B$_{\rm LoS}$ in area ROI$_2$ (black). The red curve is the sixth EMD mode with a periodicity of $50.6\pm5.8$~min. (c) The detected 50-min oscillations of average magnetic field strength in areas ROI$_{1}$--ROI$_{2}$. 
	}
	\label{fig:50min}
\end{figure*}

To clarify whether the 50-min photospheric oscillation is local, a period map \citep{2007SoPh..241..397N, 2010A&A...513A..27C} of the temporal variation of the average B$_{\rm LoS}$ in the analyzed active region is calculated from 12:45 to 17:00~UT. As in the search for the coronal oscillations, three methods are used for the time domain analysis, and only periodicity confirmed by at least two methods is accepted.
As shown in Figure~\ref{fig:50min}a, the periodicity within $45-55$~min is localized in 2 areas, including ROI$_1$. The detrended time signal of mean B$_{\rm LoS}$ in area ROI$_2$, see panel b, shows a very clear oscillation with a period of $50.6\pm5.8$~min, demonstrated by the 6th EMD mode (red curve). In panel c, the identified EMD modes exhibiting 50-min periodicity in 2 areas are lined up in parallel. They are out of phase, provided that a Pearson cross-correlation coefficient between them is 0.46, respectively. This indicates they do not share the same source.

The 50-min periodicity or its harmonics was not found in the brightness in the oscillating loop (including the loop footpoints) in any UV--EUV bandpasses of AIA, nor in the white light intensity near the loop footpoints, observed with HMI. Furthermore, a Doppler shift oscillation of 50-min is not found in the HMI Dopplergram data.
The H$\alpha$ spectroscopic data by CHASE, which however only last for 20~min each hour, show ROI$_1$ in the chromosphere contains dark fibrils and other bright small-scale structures(Fig.~\ref{fig:spectrum}c). The estimated H$\alpha$ Doppler velocity in the target region does not show any periodicity.

Fig~\ref{fig:spectrum}g--h shows the average power spectra of HMI B$_{\rm LoS}$, continuum intensity, Doppler velocity, and CHASE H$\alpha$ Doppler shift in ROI and ROI$_1$ over the analysis duration. The spectra exhibit a power-law distribution, with a broadband peak around 3.33~mHz (5~min; dotted lines) in the HMI intensity and Doppler velocity. Although an enhancement is seen near 5 min, the power at high frequencies ($>2$~mHz) remains lower than that at low frequencies, indicating the dominance of long-timescale motions in the lower atmosphere. For example, in the average spectrum of HMI continuum intensity in the ROI, the integrated power below 2~mHz is about 8 times higher than that above 2~mHz. For HMI B$_{\rm LoS}$, the ratio is 5.6. In contrast, for HMI Doppler velocity, the ratio is 0.66, but increases to 453 when the signals are not detrended.

\section{Summary and Discussion\label{sec:discussion}}
Our results demonstrate that a 50-min decayless kink oscillation of a coronal loop and a 50-min oscillation of the LoS component of the photospheric magnetic field near the loop footpoint are co-temporal. The cross-correlation coefficient between the coronal and photospheric signals is 0.73, suggesting the co-spatial and co-temporal signals are correlated. The apparent modulation of the photospheric oscillation period does not appear in the coronal signal. This may be attributed to the effect of noise and the small number of detected oscillation cycles. The 50-min kink oscillation cannot be the natural harmonic of the loop, as it would require the kink speed to have an unrealistically low value, about the coronal sound speed. 
Therefore, we conclude that the coronal 50-min oscillation is a driven solution, highly likely produced by the photospheric oscillation. The possibility of converting a compressive photospheric oscillation into a kink oscillation is consistent with previous findings \citep[e.g.,][]{2019A&A...625A.144R}.
In the same loop, another, 5-min kink oscillation is detected. If it were the natural oscillation on the second parallel harmonic, it would require a realistic kink speed of about 1000~km/s. This estimation is consistent with the majority of previous detections of decayless kink oscillations \citep{2012ApJ...751L..27W, 2013A&A...560A.107A, 2015A&A...583A.136A, 2023NatSR..1312963Z}. 
In the case considered, since there is no periodic oscillation of loop brightness, which could be a sign of vertical polarization, the 5-min eigenmode is very unlikely to be excited by longitudinal flows (such as p-modes) along the loop \citep{ 2017A&A...606A.120K}. Given consistency with previous decayless oscillations, the decayless 5-min natural oscillation of coronal loops could be produced by broadband low-frequency photospheric motions through the \lq\lq violin\rq\rq\ mechanism \citep{2016A&A...591L...5N} which requires the external driver to be quasi-stationary on the period of the natural oscillation. Moreover, this study excluded the possibility of a narrowband 5-min driver. This mechanism is supported by recent MHD simulations \citep{2020ApJ...897L..35K,2023ApJ...955...73G}. Decisive evidence for the mechanism for the detected 5-min kink oscillations requires two LoS observations, but there lack of an additional high-resolution observation of the targeted region. We suggest that the loop shows simultaneously an eigen-oscillation and a driven oscillation, which is consistent with the behaviour of a non-resonantly driven oscillator. 

\begin{figure*}
	\centering
	\includegraphics[width=\linewidth]{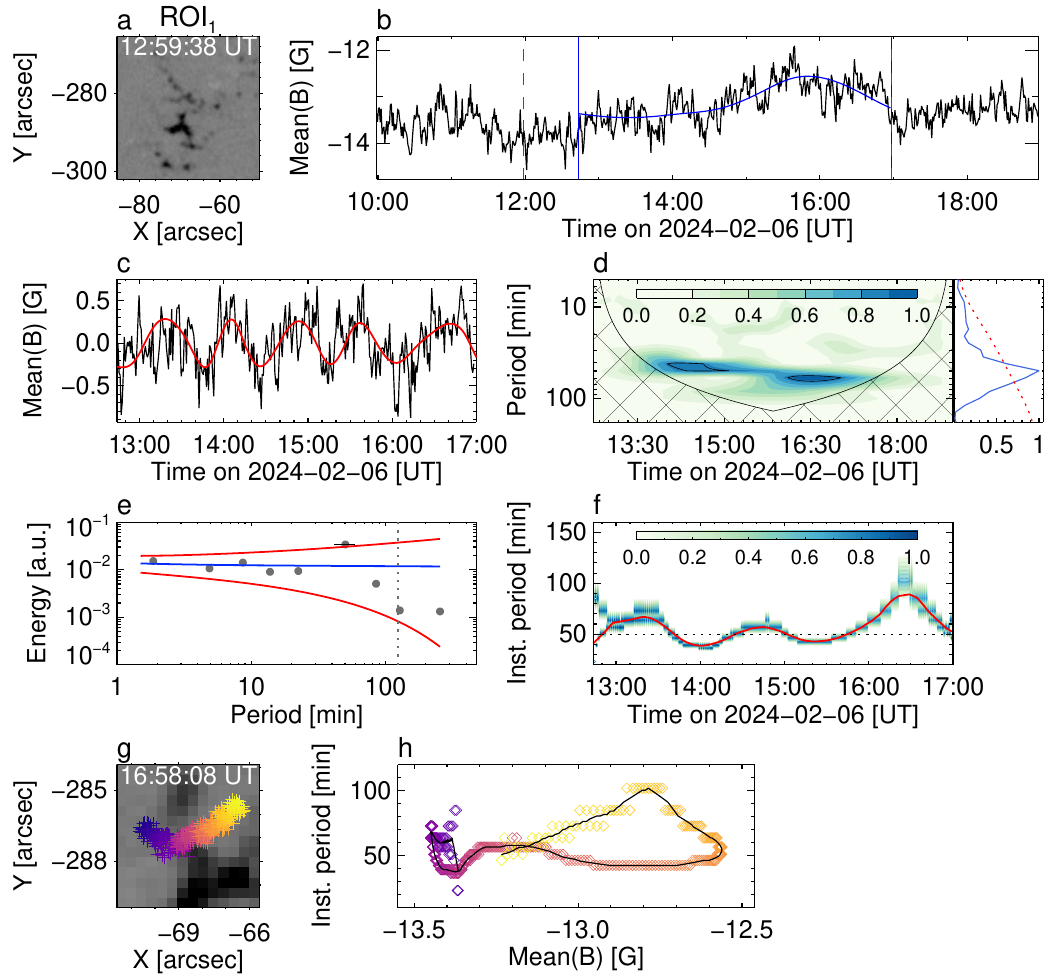}
	\caption{Temporal variation of mean B$_{\rm LoS}$ in ROI$_1$. (a) The footpoint region (ROI$_1$) of the loop of interest is seen in the HMI LoS magnetogram. (b) Time series of average LoS magnetic field in ROI$_1$. The red dashed lines indicate the time duration of the observed 50-min kink oscillations. The blue lines indicate the selected time interval where the 50-min periodicity is detected by EMD analysis. The blue curve is the trend of the selected signals. (c) The detrended time series (black) is overlaid with the detected EMD mode 6 (red). (d) Wavelet spectrum of the mean magnetic field in ROI$_1$ from 12:45~UT to 19:00~UT. The global normalized spectrum (blue) and the 95\% confidence level (red) are shown on the right of the wavelet spectrum. (e) EMD energy spectrum of the selected signals. The vertical dotted line indicates the threshold of detectable periodicity. (f) Hilbert spectrum of the EMD mode 6 shows the instantaneous period's time evolution. The red curve represents the values with maximum power. The dotted line indicates 50-min periodicity. (g) The trajectory of the centroid position of the loop footpoint. The color scheme from dark purple to yellow indicates the elapsed time since 12:00~UT. (h) Instantaneous period extracted from mode 6 as a function of the mean magnetic field. The color scheme is the same as in panel~(j). \lq\lq Inst.\rq\rq\ means instantaneous and \lq\lq a.u.\rq\rq\ stands for arbitrary units.
	}
	\label{fig:meanB}
\end{figure*}

Of interest is the nature of the 50-min oscillation of the photospheric magnetic field. 
In this work, the 50-min oscillations of the field are spatially localized, as they are situated in discrete regions covering different types of magnetic elements. Moreover, the oscillatory signals in different locations are out of phase. Hence, the global $g$-mode leakage along magnetic flux tubes \citep{2011A&A...533A.116Y} is not likely to explain the observations. The interpretation in terms of natural global oscillations of a sunspot \citep{2008AstBu..63..169S} does not seem to work either, as the eigenmode of the sunspot with 1--2~kG in our work is 30--110~min, but the target magnetic pore (in ROI$_1$) should be static as its absolute magnetic field strength of 400~G is lower than the threshold 800~G allowing for those oscillations in that model. Furthermore, the dependence of the instantaneous period on magnetic field strength extracted from mode 6 (see Figure~\ref{fig:meanB}h) is not a power-law distribution, which disagrees with the vortex shedding mechanism. The analysis of supergranulation is outside the scope of this study, as it is unclear how it could produce a narrowband oscillatory signal with the observed period. 
Furthermore, the lack of 50-min periodicity in the UV, EUV or white light intensity in/near the oscillating loop during the observation period excludes the driver associated with cyclic cooling--heating processes \citep{2016ApJ...827...39K}. More accurate determination of temperature variation can use the changes of line ratio or differential emission measures \citep[e.g.,][]{2020A&A...638A..89G}.
In general, the variation of the LoS component of the field can be caused by a kink or sausage oscillation of a photospheric flux tube \citep[e.g.,][]{2024A&A...688A...2J}. An earlier study by \citet{2005ApJS..156..265C} showed that transverse motions of magnetic bright points can excite kink waves with periods of up to several hundred minutes. A self-consistent radiative MHD simulation with Bifrost further demonstrated that convection and granulation buffeting dominated by low frequencies can trigger ubiquitous kink oscillations \citep{2021A&A...647A..81K}. However, the mechanism responsible for the periodicity detected in our work is unknown. Thus, the nature of occasionally occurring long-period photospheric oscillatory processes should be addressed in a future study.

How do the long-period photospheric oscillations propagate to the corona?
Magnetoacoustic waves generated in the photosphere can propagate up to the corona when their frequencies are higher than the cut-off frequency. In particular, some theoretical models show that the cutoff period of the kink wave is about 8--12~min \citep{2017ApJ...840...26L, 2023A&A...672A.105P}, which is significantly shorter than the detected 50-min periodicity. The magnetic field inclination could decrease the cut-off frequency for MHD waves. For example, for kink waves guided by a magnetic flux tube, the inclination of the magnetic field decreases the gradient of the kink speed along the field, caused by density stratification. This effect requires further investigation. 
The transmission of Alfv\'en waves also depends on the Alfv\'en speed ratio between layers, the ratio of 1 gives full transmission, i.e., there is no Alfv\'enic cutoff \citep[e.g.,][]{2023ApJ...954...45C}.
Moreover, in some low-atmospheric models, e.g., in an isothermal atmosphere, kink waves have been shown to have no cutoff \citep{2013ApJ...774..121L}. Additionally, wave attenuation primarily manifests as a reduction in wave amplitude, as evidenced by simulations by \citet{2023A&A...672A.105P}, where the 2000-s-period kink waves propagate to the corona at 2~km/s. In our cases, the velocity amplitude of the detected 50-min coronal oscillations is around 1~km/s, which matches the expectation. Furthermore, a wave with a period lower than the cutoff value can reach higher layers of the atmosphere because of the evanescence or tunnelling. The non-propagating nature of the detected coronal oscillation seems to be consistent with the evanescent behaviour. Those scenarios require thorough modelling which is beyond the scope of this work. Despite the lack of a solid theoretical understanding of the source of the 50-min photospheric signal, we gained evidence of the appearance of this signal in the corona. 
Further statistical studies and modeling are needed to establish the physical details of the connection, including the chromosphere's mediating role.

\begin{figure*}
	\centering
	\includegraphics[width=\linewidth]{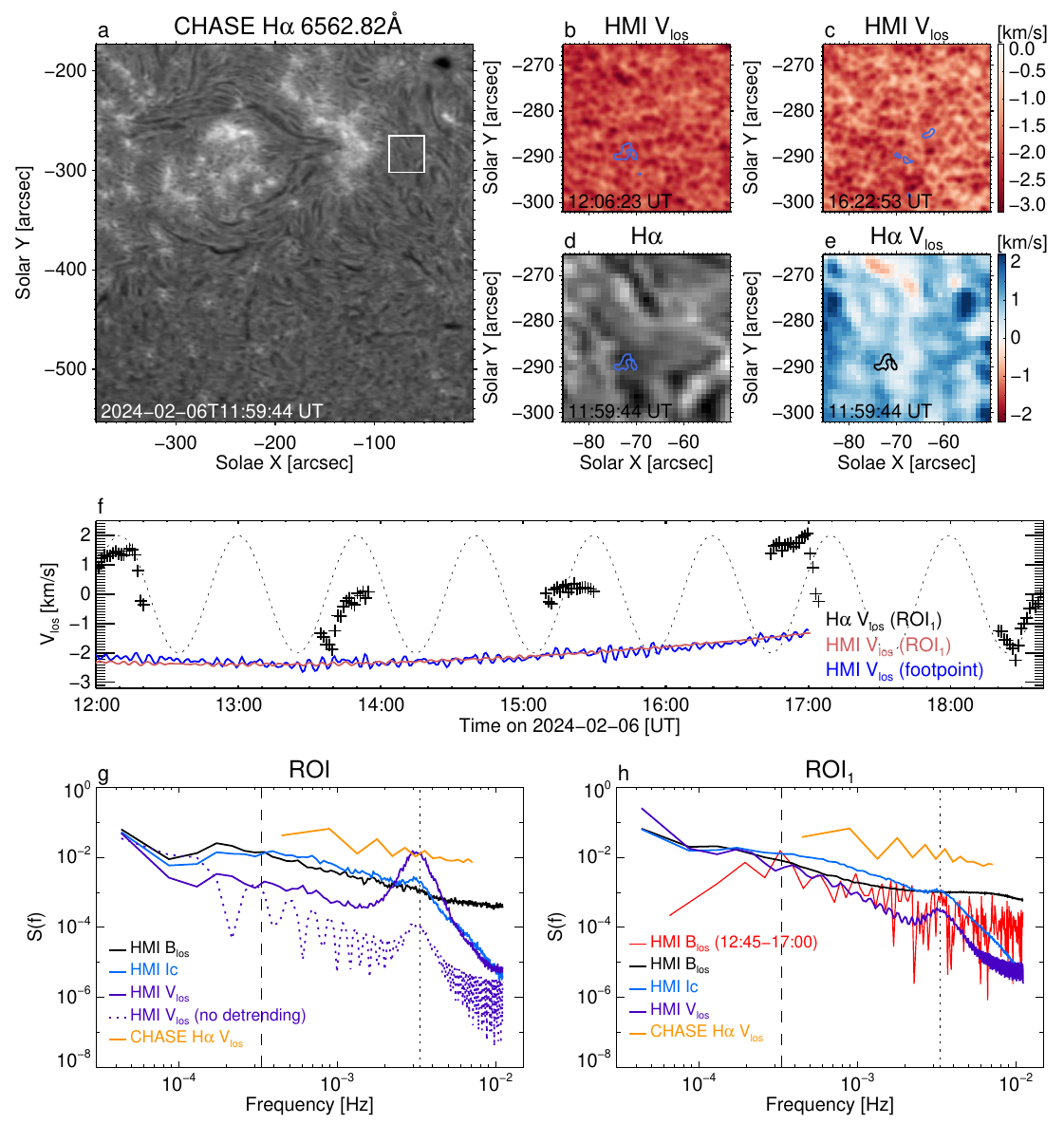}
	\caption{
    The Doppler velocity in the photosphere and chromosphere, and average power spectra from HMI data and CHASE Doppler shift. The CHASE Doppler velocity is derived from the H$\alpha$ Dopplergram after subtracting the large-scale Doppler velocity including differential rotation. (a) CHASE H$\alpha$ 6562.82\,\AA\ image of ROI. The white box marks ROI$_1$. (b--c) HMI Dopplergrams in ROI$_1$, overplotted with contours of -250~G in HMI B$_{\rm LoS}$ at the corresponding time. (d--e) H$\alpha$ 6562.82\,\AA\ image (d) and corresponding Doppler velocity map (e) in ROI$_1$. The magnetic pore is outlined by the blue and black -250~G contours of HMI B$_{\rm LoS}$. (f) Time series of average Doppler velocity in ROI$_1$, with black for H$\alpha$ LoS velocity, red for HMI Dopplergrams and blue for HMI Dopplergrams within the magnetic pore (outlined by the blue contour in panel (b)). The dotted sine curve illustrates a 50-min periodicity for comparison. (g--h) Average power spectrum of HMI B$_{\rm LoS}$ (12:00--17:00~UT), HMI continuum intensity (Ic, 12:00--17:00~UT), HMI Dopplergrams (V$_{los}$, 12:00--17:00~UT), CHASE H$\alpha$ Doppler velocity (12:00--12:20~UT) in ROI (g) and ROI$_{1}$ (h). The red curve in panel (h) shows the power spectrum of average HMI B$_{\rm LoS}$ in ROI$_{1}$ from 12:45--17:00~UT (see the black curve in Fig.~\ref{fig:meanB}c). The dotted and dashed lines mark the 5~min and 50~min timescales, respectively. Here, input signals are normalized by their standard deviation to facilitate comparison across different proxies.
	}
	\label{fig:spectrum}
\end{figure*}

Last, we discuss the potential role of low-frequency photospheric motions.
The spectrum of lower-atmospheric dynamics is characterized by power-law functions \citep{2010ApJ...710.1857M,2015ApJ...798....1I, 2016A&A...592A.153K}, i.e., colored noise. Although the spectra are usually segmented and often show a 5-min peak, the low-frequency domain contains substantial integrated power. For example, in our ROI the integrated power below 2 mHz exceeds that above 2 mHz by a factor of $\sim$8 for the HMI continuum intensity, and by $\sim$5.6 for HMI B$_{\rm LoS}$ (Fig.~\ref{fig:spectrum}g). In oscillation analysis, long-term variations are commonly removed as background noise, yet after detrending the power-law form remains, highlighting the stochastic nature of atmospheric motions.
Thus, the solar photosphere acts as a reservoir of abundant, energetically significant low-frequency motions. 
Photospheric motions in this part of the spectrum are the key element of coronal heating mechanisms based on nanoflares \citep[e.g.,][]{1988ApJ...330..474P}. Similar power-law spectra are found in the corona, both in waves \citep{2025ApJ...986L...6M} and in intensity fluctuations \citep[e.g.,][]{2008ApJ...677L.137B}. The presence of a common $1/f$-like distribution in the photosphere and corona \citep{2007ApJ...657L.121M} suggests a possible connection, with numerical simulations indicating that the coronal spectrum may reflect both the chromospheric input and partial reflections at the transition region \citep{2012ApJ...750L..33V}. Meanwhile, simulations by \citet{2025ApJ...979..236M} indicate that coronal waves excited solely by p-modes yield a flat spectrum in the low-frequency domain, inconsistent with the observed coronal power laws \citep{2019NatAs...3..223M}. Furthermore, the hypothesis of the photospheric 5-min oscillation leakage into the corona is inconsistent with the lack of a 5-min peak in the statistics of kink oscillation amplitude \citep[e.g.,][]{2016A&A...591L...5N}. This contrast implied an additional contribution from photospheric low-frequency motions.

In the context of coronal heating, attention has largely focused on short-period waves, particularly those associated with p-modes \citep[e.g.,][]{2007Sci...318.1574D,2009A&A...497..525H}. Our observations, however, highlight the potential importance of the highly energetic low-frequency part of the photospheric spectrum as a source of energy supply. Specifically, we have traced a 50-min photospheric oscillation up to the corona, where it appears as an externally driven decayless kink oscillation of a coronal loop. This result demonstrates that the low-frequency component of the photospheric spectrum can reach the corona, providing an observationally established transport channel that has not been previously identified. 
In terms of energy budget, we emphasize that low-frequency oscillations themselves are unlikely to heat the corona directly. Instead, their role is probably indirect and multifaceted. For example, such motions can interact with coronal loops through the self-oscillatory (\lq\lq violin\rq\rq) mechanism, thereby sustaining decayless kink oscillations at the loop's natural frequencies \citep{2023NatCo..14.5298Z}, which are known to dissipate efficiently. They may also excite non-resonant kink waves, as observed in the present case, which could channel energy into turbulence, ultimately leading to dissipation. In addition, low-frequency random motions can tangle magnetic fields on long timescales, providing favorable conditions for reconnection-driven nanoflare heating events \citep{1988ApJ...330..474P,2015RSPTA.37340256K}. While the precise dissipation mechanism remains uncertain, our finding suggests that low-frequency photospheric motions may play a broader role in coronal energy transport than previously appreciated. Confirming this exploratory implication will require further observational statistics and theoretical modeling. 

In the present case, we track a 50-min periodicity simultaneously in the photospheric B$_{\rm LoS}$ at a loop footpoint and in the displacement of the overlying coronal loop. To our knowledge, this is the first observational evidence of such a long-period oscillation connection between the photosphere and corona. While the exact transmission path and dissipation mechanism remain uncertain, we suggest that our results may indicate that low-frequency spectral components of photospheric motions can reach coronal heights.

\begin{acknowledgments}
The following fundings are gratefully acknowledged:
China Scholarship Council-University of Warwick joint scholarship (S.Z. during her PhD studies), the Latvian Council of Science Project No. lzp2022/1-0017 (D.Y.K. and V.M.N.), and the STFC consolidated grant ST/X000915/1 (D.Y.K).
S.Z. expresses sincere thanks to the CHASE instrument team, especially Mr. Rao Shihao, for their timely assistance with data preprocessing and for kindly sharing analysis methods and code related to large-scale Doppler velocity.
For the purpose of open access, the author has applied a Creative Commons Attribution (CC BY) licence to any author-accepted manuscript version
arising.
\end{acknowledgments}



%

\vspace{5mm}
\facilities{SDO/AIA, SDO/HMI}


\software{Astropy \citep{2013A&A...558A..33A,2018AJ....156..123A},  
          WOW \citep{2023A&A...670A..66A}, 
          SSWIDL \citep{1998SoPh..182..497F}
          }



\bibliography{50min}{}

\begin{thebibliography}{}
\expandafter\ifx\csname natexlab\endcsname\relax\def\natexlab#1{#1}\fi
\providecommand{\url}[1]{\href{#1}{#1}}
\providecommand{\dodoi}[1]{doi:~\href{http://doi.org/#1}{\nolinkurl{#1}}}
\providecommand{\doeprint}[1]{\href{http://ascl.net/#1}{\nolinkurl{http://ascl.net/#1}}}
\providecommand{\doarXiv}[1]{\href{https://arxiv.org/abs/#1}{\nolinkurl{https://arxiv.org/abs/#1}}}

\bibitem[{{Abramov-maximov} {et~al.}(2013){Abramov-maximov}, {Efremov}, {Parfinenko}, {Solov'ev}, \& {Shibasaki}}]{2013PASJ...65S..12A}
{Abramov-maximov}, V.~E., {Efremov}, V.~I., {Parfinenko}, L.~D., {Solov'ev}, A.~A., \& {Shibasaki}, K. 2013, \pasj, 65, S12, \dodoi{10.1093/pasj/65.sp1.S12}

\bibitem[{{Anfinogentov} \& {Nakariakov}(2016)}]{2016SoPh..291.3251A}
{Anfinogentov}, S., \& {Nakariakov}, V.~M. 2016, \solphys, 291, 3251, \dodoi{10.1007/s11207-016-1013-z}

\bibitem[{{Anfinogentov} {et~al.}(2013){Anfinogentov}, {Nistic{\`o}}, \& {Nakariakov}}]{2013A&A...560A.107A}
{Anfinogentov}, S., {Nistic{\`o}}, G., \& {Nakariakov}, V.~M. 2013, \aap, 560, A107, \dodoi{10.1051/0004-6361/201322094}

\bibitem[{{Anfinogentov} {et~al.}(2015){Anfinogentov}, {Nakariakov}, \& {Nistic{\`o}}}]{2015A&A...583A.136A}
{Anfinogentov}, S.~A., {Nakariakov}, V.~M., \& {Nistic{\`o}}, G. 2015, \aap, 583, A136, \dodoi{10.1051/0004-6361/201526195}

\bibitem[{{Anfinogentov} {et~al.}(2022){Anfinogentov}, {Antolin}, {Inglis}, {Kolotkov}, {Kupriyanova}, {McLaughlin}, {Nistic{\`o}}, {Pascoe}, {Krishna Prasad}, \& {Yuan}}]{2022SSRv..218....9A}
{Anfinogentov}, S.~A., {Antolin}, P., {Inglis}, A.~R., {et~al.} 2022, \ssr, 218, 9, \dodoi{10.1007/s11214-021-00869-w}

\bibitem[{{Appourchaux} {et~al.}(2010){Appourchaux}, {Belkacem}, {Broomhall}, {Chaplin}, {Gough}, {Houdek}, {Provost}, {Baudin}, {Boumier}, {Elsworth}, {Garc{\'\i}a}, {Andersen}, {Finsterle}, {Fr{\"o}hlich}, {Gabriel}, {Grec}, {Jim{\'e}nez}, {Kosovichev}, {Sekii}, {Toutain}, \& {Turck-Chi{\`e}ze}}]{2010A&ARv..18..197A}
{Appourchaux}, T., {Belkacem}, K., {Broomhall}, A.~M., {et~al.} 2010, \aapr, 18, 197, \dodoi{10.1007/s00159-009-0027-z}

\bibitem[{{Astropy Collaboration} {et~al.}(2013){Astropy Collaboration}, {Robitaille}, {Tollerud}, {Greenfield}, {Droettboom}, {Bray}, {Aldcroft}, {Davis}, {Ginsburg}, {Price-Whelan}, {Kerzendorf}, {Conley}, {Crighton}, {Barbary}, {Muna}, {Ferguson}, {Grollier}, {Parikh}, {Nair}, {Unther}, {Deil}, {Woillez}, {Conseil}, {Kramer}, {Turner}, {Singer}, {Fox}, {Weaver}, {Zabalza}, {Edwards}, {Azalee Bostroem}, {Burke}, {Casey}, {Crawford}, {Dencheva}, {Ely}, {Jenness}, {Labrie}, {Lim}, {Pierfederici}, {Pontzen}, {Ptak}, {Refsdal}, {Servillat}, \& {Streicher}}]{2013A&A...558A..33A}
{Astropy Collaboration}, {Robitaille}, T.~P., {Tollerud}, E.~J., {et~al.} 2013, \aap, 558, A33, \dodoi{10.1051/0004-6361/201322068}

\bibitem[{{Astropy Collaboration} {et~al.}(2018){Astropy Collaboration}, {Price-Whelan}, {Sip{\H{o}}cz}, {G{\"u}nther}, {Lim}, {Crawford}, {Conseil}, {Shupe}, {Craig}, {Dencheva}, {Ginsburg}, {VanderPlas}, {Bradley}, {P{\'e}rez-Su{\'a}rez}, {de Val-Borro}, {Aldcroft}, {Cruz}, {Robitaille}, {Tollerud}, {Ardelean}, {Babej}, {Bach}, {Bachetti}, {Bakanov}, {Bamford}, {Barentsen}, {Barmby}, {Baumbach}, {Berry}, {Biscani}, {Boquien}, {Bostroem}, {Bouma}, {Brammer}, {Bray}, {Breytenbach}, {Buddelmeijer}, {Burke}, {Calderone}, {Cano Rodr{\'\i}guez}, {Cara}, {Cardoso}, {Cheedella}, {Copin}, {Corrales}, {Crichton}, {D'Avella}, {Deil}, {Depagne}, {Dietrich}, {Donath}, {Droettboom}, {Earl}, {Erben}, {Fabbro}, {Ferreira}, {Finethy}, {Fox}, {Garrison}, {Gibbons}, {Goldstein}, {Gommers}, {Greco}, {Greenfield}, {Groener}, {Grollier}, {Hagen}, {Hirst}, {Homeier}, {Horton}, {Hosseinzadeh}, {Hu}, {Hunkeler}, {Ivezi{\'c}}, {Jain}, {Jenness}, {Kanarek}, {Kendrew}, {Kern}, {Kerzendorf}, {Khvalko}, {King}, {Kirkby}, {Kulkarni},
  {Kumar}, {Lee}, {Lenz}, {Littlefair}, {Ma}, {Macleod}, {Mastropietro}, {McCully}, {Montagnac}, {Morris}, {Mueller}, {Mumford}, {Muna}, {Murphy}, {Nelson}, {Nguyen}, {Ninan}, {N{\"o}the}, {Ogaz}, {Oh}, {Parejko}, {Parley}, {Pascual}, {Patil}, {Patil}, {Plunkett}, {Prochaska}, {Rastogi}, {Reddy Janga}, {Sabater}, {Sakurikar}, {Seifert}, {Sherbert}, {Sherwood-Taylor}, {Shih}, {Sick}, {Silbiger}, {Singanamalla}, {Singer}, {Sladen}, {Sooley}, {Sornarajah}, {Streicher}, {Teuben}, {Thomas}, {Tremblay}, {Turner}, {Terr{\'o}n}, {van Kerkwijk}, {de la Vega}, {Watkins}, {Weaver}, {Whitmore}, {Woillez}, {Zabalza}, \& {Astropy Contributors}}]{2018AJ....156..123A}
{Astropy Collaboration}, {Price-Whelan}, A.~M., {Sip{\H{o}}cz}, B.~M., {et~al.} 2018, \aj, 156, 123, \dodoi{10.3847/1538-3881/aabc4f}

\bibitem[{{Auch{\`e}re} {et~al.}(2023){Auch{\`e}re}, {Soubri{\'e}}, {Pelouze}, \& {Buchlin}}]{2023A&A...670A..66A}
{Auch{\`e}re}, F., {Soubri{\'e}}, E., {Pelouze}, G., \& {Buchlin}, {\'E}. 2023, \aap, 670, A66, \dodoi{10.1051/0004-6361/202245345}

\bibitem[{{Banerjee} {et~al.}(2021){Banerjee}, {Krishna Prasad}, {Pant}, {McLaughlin}, {Antolin}, {Magyar}, {Ofman}, {Tian}, {Van Doorsselaere}, {De Moortel}, \& {Wang}}]{2021SSRv..217...76B}
{Banerjee}, D., {Krishna Prasad}, S., {Pant}, V., {et~al.} 2021, \ssr, 217, 76, \dodoi{10.1007/s11214-021-00849-0}

\bibitem[{{Bemporad} {et~al.}(2008){Bemporad}, {Matthaeus}, \& {Poletto}}]{2008ApJ...677L.137B}
{Bemporad}, A., {Matthaeus}, W.~H., \& {Poletto}, G. 2008, \apjl, 677, L137, \dodoi{10.1086/588093}

\bibitem[{{Bonet} {et~al.}(2010){Bonet}, {M{\'a}rquez}, {S{\'a}nchez Almeida}, {Palacios}, {Mart{\'\i}nez Pillet}, {Solanki}, {del Toro Iniesta}, {Domingo}, {Berkefeld}, {Schmidt}, {Gandorfer}, {Barthol}, \& {Kn{\"o}lker}}]{2010ApJ...723L.139B}
{Bonet}, J.~A., {M{\'a}rquez}, I., {S{\'a}nchez Almeida}, J., {et~al.} 2010, \apjl, 723, L139, \dodoi{10.1088/2041-8205/723/2/L139}

\bibitem[{{Chae} \& {Lee}(2023)}]{2023ApJ...954...45C}
{Chae}, J., \& {Lee}, K.-S. 2023, \apj, 954, 45, \dodoi{10.3847/1538-4357/ace771}

\bibitem[{{Chelpanov} \& {Kobanov}(2022)}]{2022SoPh..297..154C}
{Chelpanov}, A., \& {Kobanov}, N. 2022, \solphys, 297, 154, \dodoi{10.1007/s11207-022-02092-4}

\bibitem[{{Chitta} {et~al.}(2012){Chitta}, {van Ballegooijen}, {Rouppe van der Voort}, {DeLuca}, \& {Kariyappa}}]{2012ApJ...752...48C}
{Chitta}, L.~P., {van Ballegooijen}, A.~A., {Rouppe van der Voort}, L., {DeLuca}, E.~E., \& {Kariyappa}, R. 2012, \apj, 752, 48, \dodoi{10.1088/0004-637X/752/1/48}

\bibitem[{{Chorley} {et~al.}(2010){Chorley}, {Hnat}, {Nakariakov}, {Inglis}, \& {Bakunina}}]{2010A&A...513A..27C}
{Chorley}, N., {Hnat}, B., {Nakariakov}, V.~M., {Inglis}, A.~R., \& {Bakunina}, I.~A. 2010, \aap, 513, A27, \dodoi{10.1051/0004-6361/200913683}

\bibitem[{{Cranmer} \& {van Ballegooijen}(2005)}]{2005ApJS..156..265C}
{Cranmer}, S.~R., \& {van Ballegooijen}, A.~A. 2005, \apjs, 156, 265, \dodoi{10.1086/426507}

\bibitem[{{De Moortel} \& {Browning}(2015)}]{2015RSPTA.37340269D}
{De Moortel}, I., \& {Browning}, P. 2015, Philosophical Transactions of the Royal Society of London Series A, 373, 20140269, \dodoi{10.1098/rsta.2014.0269}

\bibitem[{{De Pontieu} {et~al.}(2007){De Pontieu}, {McIntosh}, {Carlsson}, {Hansteen}, {Tarbell}, {Schrijver}, {Title}, {Shine}, {Tsuneta}, {Katsukawa}, {Ichimoto}, {Suematsu}, {Shimizu}, \& {Nagata}}]{2007Sci...318.1574D}
{De Pontieu}, B., {McIntosh}, S.~W., {Carlsson}, M., {et~al.} 2007, Science, 318, 1574, \dodoi{10.1126/science.1151747}

\bibitem[{{Dumbadze} {et~al.}(2021){Dumbadze}, {Shergelashvili}, {Poedts}, {Zaqarashvili}, {Khodachenko}, \& {De Causmaecker}}]{2021A&A...653A..39D}
{Dumbadze}, G., {Shergelashvili}, B.~M., {Poedts}, S., {et~al.} 2021, \aap, 653, A39, \dodoi{10.1051/0004-6361/202038124}

\bibitem[{{Dumbadze} {et~al.}(2017){Dumbadze}, {Shergelashvili}, {Kukhianidze}, {Ramishvili}, {Zaqarashvili}, {Khodachenko}, {Gurgenashvili}, {Poedts}, \& {De Causmaecker}}]{2017A&A...597A..93D}
{Dumbadze}, G., {Shergelashvili}, B.~M., {Kukhianidze}, V., {et~al.} 2017, \aap, 597, A93, \dodoi{10.1051/0004-6361/201628213}

\bibitem[{{Efremov} {et~al.}(2007){Efremov}, {Parfinenko}, \& {Solov'ev}}]{2007ARep...51..401E}
{Efremov}, V.~I., {Parfinenko}, L.~D., \& {Solov'ev}, A.~A. 2007, Astronomy Reports, 51, 401, \dodoi{10.1134/S106377290705006X}

\bibitem[{{Efremov} {et~al.}(2009){Efremov}, {Parfinenko}, \& {Soloviev}}]{2009CosRe..47..279E}
{Efremov}, V.~I., {Parfinenko}, L.~D., \& {Soloviev}, A.~A. 2009, Cosmic Research, 47, 279, \dodoi{10.1134/S0010952509040030}

\bibitem[{{Emonet} {et~al.}(2001){Emonet}, {Moreno-Insertis}, \& {Rast}}]{2001ApJ...549.1212E}
{Emonet}, T., {Moreno-Insertis}, F., \& {Rast}, M.~P. 2001, \apj, 549, 1212, \dodoi{10.1086/319469}

\bibitem[{{Freeland} \& {Handy}(1998)}]{1998SoPh..182..497F}
{Freeland}, S.~L., \& {Handy}, B.~N. 1998, \solphys, 182, 497, \dodoi{10.1023/A:1005038224881}

\bibitem[{{Gao} {et~al.}(2023){Gao}, {Guo}, {Van Doorsselaere}, {Tian}, \& {Skirvin}}]{2023ApJ...955...73G}
{Gao}, Y., {Guo}, M., {Van Doorsselaere}, T., {Tian}, H., \& {Skirvin}, S.~J. 2023, \apj, 955, 73, \dodoi{10.3847/1538-4357/acf454}

\bibitem[{{Gelfreikh} {et~al.}(2006){Gelfreikh}, {Nagovitsyn}, \& {Nagovitsyna}}]{2006PASJ...58...29G}
{Gelfreikh}, G.~B., {Nagovitsyn}, Y.~A., \& {Nagovitsyna}, E.~Y. 2006, \pasj, 58, 29, \dodoi{10.1093/pasj/58.1.29}

\bibitem[{{Goddard} \& {Nistic{\`o}}(2020)}]{2020A&A...638A..89G}
{Goddard}, C.~R., \& {Nistic{\`o}}, G. 2020, \aap, 638, A89, \dodoi{10.1051/0004-6361/202037467}

\bibitem[{{Gri{\~n}{\'o}n-Mar{\'\i}n} {et~al.}(2020){Gri{\~n}{\'o}n-Mar{\'\i}n}, {Pastor Yabar}, {Socas-Navarro}, \& {Centeno}}]{2020A&A...635A..64G}
{Gri{\~n}{\'o}n-Mar{\'\i}n}, A.~B., {Pastor Yabar}, A., {Socas-Navarro}, H., \& {Centeno}, R. 2020, \aap, 635, A64, \dodoi{10.1051/0004-6361/201936589}

\bibitem[{{He} {et~al.}(2009){He}, {Tu}, {Marsch}, {Guo}, {Yao}, \& {Tian}}]{2009A&A...497..525H}
{He}, J.~S., {Tu}, C.~Y., {Marsch}, E., {et~al.} 2009, \aap, 497, 525, \dodoi{10.1051/0004-6361/200810777}

\bibitem[{{Huang} \& {Wu}(2008)}]{2008RvGeo..46.2006H}
{Huang}, N.~E., \& {Wu}, Z. 2008, Reviews of Geophysics, 46, RG2006, \dodoi{10.1029/2007RG000228}

\bibitem[{{Huang} {et~al.}(1998){Huang}, {Shen}, {Long}, {Wu}, {Shih}, {Zheng}, {Yen}, {Tung}, \& {Liu}}]{1998RSPSA.454..903H}
{Huang}, N.~E., {Shen}, Z., {Long}, S.~R., {et~al.} 1998, Proceedings of the Royal Society of London Series A, 454, 903, \dodoi{10.1098/rspa.1998.0193}

\bibitem[{{Ireland} {et~al.}(2015){Ireland}, {McAteer}, \& {Inglis}}]{2015ApJ...798....1I}
{Ireland}, J., {McAteer}, R.~T.~J., \& {Inglis}, A.~R. 2015, \apj, 798, 1, \dodoi{10.1088/0004-637X/798/1/1}

\bibitem[{{Jafarzadeh} {et~al.}(2024){Jafarzadeh}, {Schiavo}, {Fedun}, {Solanki}, {Stangalini}, {Calchetti}, {Verth}, {Jess}, {Grant}, {Ballai}, {Gafeira}, {Keys}, {Fleck}, {Morton}, {Browning}, {Silva}, {Appourchaux}, {Gandorfer}, {Gizon}, {Hirzberger}, {Kahil}, {Orozco Su{\'a}rez}, {Schou}, {Strecker}, {del Toro Iniesta}, {Valori}, {Volkmer}, \& {Woch}}]{2024A&A...688A...2J}
{Jafarzadeh}, S., {Schiavo}, L.~A.~C.~A., {Fedun}, V., {et~al.} 2024, \aap, 688, A2, \dodoi{10.1051/0004-6361/202449685}

\bibitem[{{Karampelas} \& {Van Doorsselaere}(2020)}]{2020ApJ...897L..35K}
{Karampelas}, K., \& {Van Doorsselaere}, T. 2020, \apjl, 897, L35, \dodoi{10.3847/2041-8213/ab9f38}

\bibitem[{{Klimchuk}(2015)}]{2015RSPTA.37340256K}
{Klimchuk}, J.~A. 2015, Philosophical Transactions of the Royal Society of London Series A, 373, 20140256, \dodoi{10.1098/rsta.2014.0256}

\bibitem[{{Kobrin} \& {Korshunov}(1972)}]{1972SoPh...25..339K}
{Kobrin}, M.~M., \& {Korshunov}, A.~I. 1972, \solphys, 25, 339, \dodoi{10.1007/BF00192332}

\bibitem[{{Kohutova} \& {Popovas}(2021)}]{2021A&A...647A..81K}
{Kohutova}, P., \& {Popovas}, A. 2021, \aap, 647, A81, \dodoi{10.1051/0004-6361/202039491}

\bibitem[{{Kohutova} \& {Verwichte}(2016)}]{2016ApJ...827...39K}
{Kohutova}, P., \& {Verwichte}, E. 2016, \apj, 827, 39, \dodoi{10.3847/0004-637X/827/1/39}

\bibitem[{{Kohutova} \& {Verwichte}(2017)}]{2017A&A...606A.120K}
---. 2017, \aap, 606, A120, \dodoi{10.1051/0004-6361/201731417}

\bibitem[{{Kolotkov} {et~al.}(2016){Kolotkov}, {Anfinogentov}, \& {Nakariakov}}]{2016A&A...592A.153K}
{Kolotkov}, D.~Y., {Anfinogentov}, S.~A., \& {Nakariakov}, V.~M. 2016, \aap, 592, A153, \dodoi{10.1051/0004-6361/201628306}

\bibitem[{{Kolotkov} {et~al.}(2017){Kolotkov}, {Smirnova}, {Strekalova}, {Riehokainen}, \& {Nakariakov}}]{2017A&A...598L...2K}
{Kolotkov}, D.~Y., {Smirnova}, V.~V., {Strekalova}, P.~V., {Riehokainen}, A., \& {Nakariakov}, V.~M. 2017, \aap, 598, L2, \dodoi{10.1051/0004-6361/201629951}

\bibitem[{{Lemen} {et~al.}(2012){Lemen}, {Title}, {Akin}, {Boerner}, {Chou}, {Drake}, {Duncan}, {Edwards}, {Friedlaender}, {Heyman}, {Hurlburt}, {Katz}, {Kushner}, {Levay}, {Lindgren}, {Mathur}, {McFeaters}, {Mitchell}, {Rehse}, {Schrijver}, {Springer}, {Stern}, {Tarbell}, {Wuelser}, {Wolfson}, {Yanari}, {Bookbinder}, {Cheimets}, {Caldwell}, {Deluca}, {Gates}, {Golub}, {Park}, {Podgorski}, {Bush}, {Scherrer}, {Gummin}, {Smith}, {Auker}, {Jerram}, {Pool}, {Soufli}, {Windt}, {Beardsley}, {Clapp}, {Lang}, \& {Waltham}}]{2012SoPh..275...17L}
{Lemen}, J.~R., {Title}, A.~M., {Akin}, D.~J., {et~al.} 2012, \solphys, 275, 17, \dodoi{10.1007/s11207-011-9776-8}

\bibitem[{{Li} {et~al.}(2020){Li}, {Antolin}, {Guo}, {Kuznetsov}, {Pascoe}, {Van Doorsselaere}, \& {Vasheghani Farahani}}]{2020SSRv..216..136L}
{Li}, B., {Antolin}, P., {Guo}, M.~Z., {et~al.} 2020, \ssr, 216, 136, \dodoi{10.1007/s11214-020-00761-z}

\bibitem[{{Li} {et~al.}(2022){Li}, {Fang}, {Li}, {Ding}, {Chen}, {Qiu}, {You}, {Yuan}, {An}, {Tao}, {Li}, {Chen}, {Liu}, {Mei}, {Yang}, {Zhang}, {Cheng}, {Chen}, {Chen}, {Gu}, {Huang}, {Liu}, {Han}, {Xin}, {Chen}, {Ni}, {Wang}, {Rao}, {Li}, {Lu}, {Wang}, {Lin}, {Jiang}, {Meng}, \& {Zhao}}]{2022SCPMA..6589602L}
{Li}, C., {Fang}, C., {Li}, Z., {et~al.} 2022, Science China Physics, Mechanics, and Astronomy, 65, 289602, \dodoi{10.1007/s11433-022-1893-3}

\bibitem[{{Liu} \& {Ofman}(2014)}]{2014SoPh..289.3233L}
{Liu}, W., \& {Ofman}, L. 2014, \solphys, 289, 3233, \dodoi{10.1007/s11207-014-0528-4}

\bibitem[{{Lopin} \& {Nagorny}(2013)}]{2013ApJ...774..121L}
{Lopin}, I., \& {Nagorny}, I. 2013, \apj, 774, 121, \dodoi{10.1088/0004-637X/774/2/121}

\bibitem[{{Lopin} \& {Nagorny}(2017)}]{2017ApJ...840...26L}
---. 2017, \apj, 840, 26, \dodoi{10.3847/1538-4357/aa6c5a}

\bibitem[{{Matsumoto} \& {Kitai}(2010)}]{2010ApJ...716L..19M}
{Matsumoto}, T., \& {Kitai}, R. 2010, \apjl, 716, L19, \dodoi{10.1088/2041-8205/716/1/L19}

\bibitem[{{Matsumoto} \& {Shibata}(2010)}]{2010ApJ...710.1857M}
{Matsumoto}, T., \& {Shibata}, K. 2010, \apj, 710, 1857, \dodoi{10.1088/0004-637X/710/2/1857}

\bibitem[{{Matthaeus} {et~al.}(2007){Matthaeus}, {Breech}, {Dmitruk}, {Bemporad}, {Poletto}, {Velli}, \& {Romoli}}]{2007ApJ...657L.121M}
{Matthaeus}, W.~H., {Breech}, B., {Dmitruk}, P., {et~al.} 2007, \apjl, 657, L121, \dodoi{10.1086/513075}

\bibitem[{{Miriyala} {et~al.}(2025){Miriyala}, {Morton}, {Khomenko}, {Antolin}, \& {Botha}}]{2025ApJ...979..236M}
{Miriyala}, H., {Morton}, R.~J., {Khomenko}, E., {Antolin}, P., \& {Botha}, G. J.~J. 2025, \apj, 979, 236, \dodoi{10.3847/1538-4357/ada26f}

\bibitem[{{Morton} \& {Soler}(2025)}]{2025ApJ...986L...6M}
{Morton}, R.~J., \& {Soler}, R. 2025, \apjl, 986, L6, \dodoi{10.3847/2041-8213/add7da}

\bibitem[{{Morton} {et~al.}(2019){Morton}, {Weberg}, \& {McLaughlin}}]{2019NatAs...3..223M}
{Morton}, R.~J., {Weberg}, M.~J., \& {McLaughlin}, J.~A. 2019, Nature Astronomy, 3, 223, \dodoi{10.1038/s41550-018-0668-9}

\bibitem[{{Nakariakov} {et~al.}(2016){Nakariakov}, {Anfinogentov}, {Nistic{\`o}}, \& {Lee}}]{2016A&A...591L...5N}
{Nakariakov}, V.~M., {Anfinogentov}, S.~A., {Nistic{\`o}}, G., \& {Lee}, D.~H. 2016, \aap, 591, L5, \dodoi{10.1051/0004-6361/201628850}

\bibitem[{{Nakariakov} \& {King}(2007)}]{2007SoPh..241..397N}
{Nakariakov}, V.~M., \& {King}, D.~B. 2007, \solphys, 241, 397, \dodoi{10.1007/s11207-007-0348-x}

\bibitem[{{Nakariakov} \& {Kolotkov}(2020)}]{2020ARA&A..58..441N}
{Nakariakov}, V.~M., \& {Kolotkov}, D.~Y. 2020, \araa, 58, 441, \dodoi{10.1146/annurev-astro-032320-042940}

\bibitem[{{Nakariakov} {et~al.}(2021){Nakariakov}, {Anfinogentov}, {Antolin}, {Jain}, {Kolotkov}, {Kupriyanova}, {Li}, {Magyar}, {Nistic{\`o}}, {Pascoe}, {Srivastava}, {Terradas}, {Vasheghani Farahani}, {Verth}, {Yuan}, \& {Zimovets}}]{2021SSRv..217...73N}
{Nakariakov}, V.~M., {Anfinogentov}, S.~A., {Antolin}, P., {et~al.} 2021, \ssr, 217, 73, \dodoi{10.1007/s11214-021-00847-2}

\bibitem[{{Nechaeva} {et~al.}(2019){Nechaeva}, {Zimovets}, {Nakariakov}, \& {Goddard}}]{2019ApJS..241...31N}
{Nechaeva}, A., {Zimovets}, I.~V., {Nakariakov}, V.~M., \& {Goddard}, C.~R. 2019, \apjs, 241, 31, \dodoi{10.3847/1538-4365/ab0e86}

\bibitem[{{Nelson} {et~al.}(2021){Nelson}, {Campbell}, \& {Mathioudakis}}]{2021A&A...654A..50N}
{Nelson}, C.~J., {Campbell}, R.~J., \& {Mathioudakis}, M. 2021, \aap, 654, A50, \dodoi{10.1051/0004-6361/202141368}

\bibitem[{{Parker}(1988)}]{1988ApJ...330..474P}
{Parker}, E.~N. 1988, \apj, 330, 474, \dodoi{10.1086/166485}

\bibitem[{{Pelouze} {et~al.}(2023){Pelouze}, {Van Doorsselaere}, {Karampelas}, {Riedl}, \& {Duckenfield}}]{2023A&A...672A.105P}
{Pelouze}, G., {Van Doorsselaere}, T., {Karampelas}, K., {Riedl}, J.~M., \& {Duckenfield}, T. 2023, \aap, 672, A105, \dodoi{10.1051/0004-6361/202245049}

\bibitem[{{Pesnell} {et~al.}(2012){Pesnell}, {Thompson}, \& {Chamberlin}}]{2012SoPh..275....3P}
{Pesnell}, W.~D., {Thompson}, B.~J., \& {Chamberlin}, P.~C. 2012, \solphys, 275, 3, \dodoi{10.1007/s11207-011-9841-3}

\bibitem[{{Pugh} {et~al.}(2017){Pugh}, {Broomhall}, \& {Nakariakov}}]{2017A&A...602A..47P}
{Pugh}, C.~E., {Broomhall}, A.~M., \& {Nakariakov}, V.~M. 2017, \aap, 602, A47, \dodoi{10.1051/0004-6361/201730595}

\bibitem[{{Rao} {et~al.}(2024){Rao}, {Li}, {Ding}, {Hong}, {Chen}, {Fang}, {Qiu}, {Li}, {Chen}, {Li}, {Hao}, {Guo}, {Cheng}, {Dai}, {Peng}, {You}, \& {Yuan}}]{2024NatAs...8.1102R}
{Rao}, S., {Li}, C., {Ding}, M., {et~al.} 2024, Nature Astronomy, 8, 1102, \dodoi{10.1038/s41550-024-02299-4}

\bibitem[{{Riedl} {et~al.}(2019){Riedl}, {Van Doorsselaere}, \& {Santamaria}}]{2019A&A...625A.144R}
{Riedl}, J.~M., {Van Doorsselaere}, T., \& {Santamaria}, I.~C. 2019, \aap, 625, A144, \dodoi{10.1051/0004-6361/201935393}

\bibitem[{{Scherrer} {et~al.}(2012){Scherrer}, {Schou}, {Bush}, {Kosovichev}, {Bogart}, {Hoeksema}, {Liu}, {Duvall}, {Zhao}, {Title}, {Schrijver}, {Tarbell}, \& {Tomczyk}}]{2012SoPh..275..207S}
{Scherrer}, P.~H., {Schou}, J., {Bush}, R.~I., {et~al.} 2012, \solphys, 275, 207, \dodoi{10.1007/s11207-011-9834-2}

\bibitem[{{Smirnova} {et~al.}(2013){Smirnova}, {Riehokainen}, {Solov'ev}, {Kallunki}, {Zhiltsov}, \& {Ryzhov}}]{2013A&A...552A..23S}
{Smirnova}, V., {Riehokainen}, A., {Solov'ev}, A., {et~al.} 2013, \aap, 552, A23, \dodoi{10.1051/0004-6361/201219600}

\bibitem[{{Solov'ev} \& {Kirichek}(2008)}]{2008AstBu..63..169S}
{Solov'ev}, A.~A., \& {Kirichek}, E.~A. 2008, Astrophysical Bulletin, 63, 169, \dodoi{10.1134/S1990341308020077}

\bibitem[{{Stangalini} {et~al.}(2013{\natexlab{a}}){Stangalini}, {Berrilli}, \& {Consolini}}]{2013A&A...559A..88S}
{Stangalini}, M., {Berrilli}, F., \& {Consolini}, G. 2013{\natexlab{a}}, \aap, 559, A88, \dodoi{10.1051/0004-6361/201322163}

\bibitem[{{Stangalini} {et~al.}(2013{\natexlab{b}}){Stangalini}, {Solanki}, {Cameron}, \& {Mart{\'\i}nez Pillet}}]{2013A&A...554A.115S}
{Stangalini}, M., {Solanki}, S.~K., {Cameron}, R., \& {Mart{\'\i}nez Pillet}, V. 2013{\natexlab{b}}, \aap, 554, A115, \dodoi{10.1051/0004-6361/201220933}

\bibitem[{{Strekalova} {et~al.}(2016){Strekalova}, {Nagovitsyn}, {Riehokainen}, \& {Smirnova}}]{2016Ge&Ae..56.1052S}
{Strekalova}, P.~V., {Nagovitsyn}, Y.~A., {Riehokainen}, A., \& {Smirnova}, V.~V. 2016, Geomagnetism and Aeronomy, 56, 1052, \dodoi{10.1134/S0016793216080211}

\bibitem[{{Torrence} \& {Compo}(1998)}]{1998BAMS...79...61T}
{Torrence}, C., \& {Compo}, G.~P. 1998, Bulletin of the American Meteorological Society, 79, 61, \dodoi{10.1175/1520-0477(1998)079<0061:APGTWA>2.0.CO;2}

\bibitem[{{Van Doorsselaere} {et~al.}(2020){Van Doorsselaere}, {Srivastava}, {Antolin}, {Magyar}, {Vasheghani Farahani}, {Tian}, {Kolotkov}, {Ofman}, {Guo}, {Arregui}, {De Moortel}, \& {Pascoe}}]{2020SSRv..216..140V}
{Van Doorsselaere}, T., {Srivastava}, A.~K., {Antolin}, P., {et~al.} 2020, \ssr, 216, 140, \dodoi{10.1007/s11214-020-00770-y}

\bibitem[{{Verdini} {et~al.}(2012){Verdini}, {Grappin}, {Pinto}, \& {Velli}}]{2012ApJ...750L..33V}
{Verdini}, A., {Grappin}, R., {Pinto}, R., \& {Velli}, M. 2012, \apjl, 750, L33, \dodoi{10.1088/2041-8205/750/2/L33}

\bibitem[{{Wang} {et~al.}(2012){Wang}, {Ofman}, {Davila}, \& {Su}}]{2012ApJ...751L..27W}
{Wang}, T., {Ofman}, L., {Davila}, J.~M., \& {Su}, Y. 2012, \apjl, 751, L27, \dodoi{10.1088/2041-8205/751/2/L27}

\bibitem[{{Wang} {et~al.}(2021){Wang}, {Ofman}, {Yuan}, {Reale}, {Kolotkov}, \& {Srivastava}}]{2021SSRv..217...34W}
{Wang}, T., {Ofman}, L., {Yuan}, D., {et~al.} 2021, \ssr, 217, 34, \dodoi{10.1007/s11214-021-00811-0}

\bibitem[{{Yu} {et~al.}(2024){Yu}, {Rao}, {Zhao}, {Li}, {Su}, {Zhao}, {Qiu}, {Ding}, {Fang}, {Li}, \& {Gan}}]{2024ApJ...968L..20Y}
{Yu}, F., {Rao}, S., {Zhao}, J., {et~al.} 2024, \apjl, 968, L20, \dodoi{10.3847/2041-8213/ad50c7}

\bibitem[{{Yuan} {et~al.}(2011){Yuan}, {Nakariakov}, {Chorley}, \& {Foullon}}]{2011A&A...533A.116Y}
{Yuan}, D., {Nakariakov}, V.~M., {Chorley}, N., \& {Foullon}, C. 2011, \aap, 533, A116, \dodoi{10.1051/0004-6361/201116933}

\bibitem[{{Zhong} {et~al.}(2021){Zhong}, {Duckenfield}, {Nakariakov}, \& {Anfinogentov}}]{2021SoPh..296..135Z}
{Zhong}, S., {Duckenfield}, T.~J., {Nakariakov}, V.~M., \& {Anfinogentov}, S.~A. 2021, \solphys, 296, 135, \dodoi{10.1007/s11207-021-01870-w}

\bibitem[{{Zhong} {et~al.}(2023{\natexlab{a}}){Zhong}, {Nakariakov}, {Kolotkov}, {Chitta}, {Antolin}, {Verbeeck}, \& {Berghmans}}]{2023NatCo..14.5298Z}
{Zhong}, S., {Nakariakov}, V.~M., {Kolotkov}, D.~Y., {et~al.} 2023{\natexlab{a}}, Nature Communications, 14, 5298, \dodoi{10.1038/s41467-023-41029-8}

\bibitem[{{Zhong} {et~al.}(2023{\natexlab{b}}){Zhong}, {Nakariakov}, {Miao}, {Fu}, \& {Yuan}}]{2023NatSR..1312963Z}
{Zhong}, S., {Nakariakov}, V.~M., {Miao}, Y., {Fu}, L., \& {Yuan}, D. 2023{\natexlab{b}}, Scientific Reports, 13, 12963, \dodoi{10.1038/s41598-023-40063-2}

\end{thebibliography}
\bibliographystyle{aasjournal}



\end{document}